\documentclass[aip,1p,eprint]{revtex4-1}
\usepackage{amssymb}
\usepackage{amsmath}
\usepackage{lipsum}
\usepackage{booktabs}
\usepackage{epstopdf}
\usepackage{dcolumn}%
\usepackage{graphicx}% http://ctan.org/pkg/graphicx
\usepackage[bookmarks=false,pdfstartview=FitH]{hyperref}
\newcommand\T{\rule{0pt}{2.6ex}}       % Top strut
\newcommand\B{\rule[-1.2ex]{0pt}{0pt}} % Bottom strut

\begin{document}


\title{First-principles DFT + \emph{GW} study of the Te antisite in CdTe}

\author{Mauricio A. Flores}
\email[]{mauricio.flores@ug.uchile.cl}
%\thanks{}
\affiliation{Departamento de F\'isica, Facultad de Ciencias, Universidad de Chile, Las Palmeras 3425, 780-0003 \~Nu\~noa, Santiago, Chile.}

\author{Eduardo Men\'endez-Proupin}
%\thanks{}
\affiliation{Departamento de F\'isica, Facultad de Ciencias, Universidad de Chile, Las Palmeras 3425, 780-0003 \~Nu\~noa, Santiago, Chile.}

\author{Walter Orellana}
\affiliation{Departamento de Ciencias F\'isicas, Universidad Andres Bello, Sazi\'e 2212, 037-0136 Santiago, Chile.}

\begin{abstract}
Formation energies, charge transitions levels, and quasiparticle defect states of the tellurium antisite $(\text{Te}_\text{Cd})$ in CdTe are addressed within the DFT$\hspace{0.05cm}+\hspace{0.05cm}$\emph{GW} formalism. We find that $(\text{Te}_\text{Cd})$ induces a (+2/0) deep level at 0.99 eV above the valence band maximum, exhibiting a negative-U effect. Moreover, the calculated zero-phonon line for the excited state of $(\text{Te}_\text{Cd})^0$ corresponds closely with the $\sim$1.1 eV band, visible in luminescence and absorption experiments. Our results differ from previous theoretical studies, mainly due to the well-known band gap error and the incorrect position of the band edges predicted by standard DFT calculations.
\end{abstract}

\date{\today}

\maketitle

\section{Introduction}
\label{Introduction}

Cadmium telluride (CdTe) is becoming an increasingly important II-VI semiconductor that can be obtained
with both $n$- and $p$-type conductivity \cite{Segall63,Khattak91,Basol87,Hofmann92}. Its main applications include room temperature $x$-ray and $\gamma$-ray detectors, medical imaging, nuclear safeguards, and thin-film solar cells \cite{Szeles04,Schlesinger01,Shah99}. CdTe has a high optical absorption coefficient and a near-ideal direct band gap of $\sim$1.5 eV at room temperature, which is optimum for solar energy conversion. However, native defects and impurities usually form compensating donors and acceptors that decrease both carrier concentration, and lifetime \cite{hage1992,fiederle1994,Krsmanovic00}. Consequently, controlled doping with Cu \cite{Kranz13,Korevaar14,Yang16_1} and Cl \cite{Metzger06,Li13} is commonly used to enhance hole density and carrier lifetime. Furthermore, deep levels may act as recombination centers that are detrimental to electron transport, thereby degrading the performance of solar cells and high-energy radiation detectors.

High resistivity of undoped CdTe has been associated with the Fermi level pinning near midgap by a native deep donor, which is usually assumed to be tellurium antisite $(\text{Te}_\text{Cd})$ or interstitial tellurium $(\text{Te}_\text{i})$ \cite{Chu01,Fiederle04,Babentsov09,Ma13}, considering that CdTe is normally grown in a Te-rich environment. However, theoretical results show that $(\text{Te}_\text{Cd})$ induces a gap level that is too shallow to pin the Fermi level close to the midgap \cite{Chu01,Fiederle98,Du08}, whereas $(\text{Te}_\text{i})$ has higher formation energy than $(\text{Te}_\text{Cd})$ in the Te-rich limit \cite{Yang14}. Moreover, the 1.1-eV band usually observed in luminescence and absorption experiments, remains as an unresolved issue. In 1968, Bryant and Webster \cite{Bryant68} associated it to the Te vacancy, but theory has not yet confirmed this.

  The theoretical description of defects and impurities in semiconductors is currently performed in the framework of the density functional theory (DFT), which reduces the many-electron problem to an effective single-electron problem. In principle, DFT provides an exact formulation to calculate ground-state properties, but it fails to predict the band gaps of semiconductors and insulators as there is no theoretical support for interpreting the eigenvalues from the Kohn-Sham equations as quasiparticle energies. Moreover, neglecting correlation effects can give qualitatively incorrect results for systems with partially filled electronic $d$ or $f$ shells. Additionally, the self-interaction error artificially raises the position of the valence band maximum (VBM) \cite{Gruneis14,Du15,Freysoldt16} and may lead to unreliable defect-level positions in the band gap. This is particularly severe for deep defect levels such as $(\text{Te}_\text{Cd})$. All these limitations involved in DFT calculations make them not reliable to evaluate defect properties, such as formation energies and charge transition levels \cite{Biswas11,Petretto15}. On the other hand, the \emph{GW} formalism \cite{Hedin65,Hybertsen85}, which describes the interaction of weakly correlated
quasiparticles by means of a nonlocal energy-dependent self-energy, can give accurate quasiparticle band structures of solids \cite{Zakharov94,Klimes14}.

Early DFT calculations of native defects in CdTe suggest that $(\text{Te}_\text{Cd})$ is stable in $(+2)$, $(+1)$, and neutral charge states \cite{Berding99,S-Huai02}. Du \emph{et al.} \cite{Du08_2} have found that (Te$_\text{Cd}$) exhibits a negative-U behavior with a (+2/0) transition level at VBM + 0.35 eV. On the other hand, Carvalho \emph{et al.} \cite{Carvalho10} using the local spin density approximation (LSDA) found no negative-U effect. More recent calculations employing hybrid functionals that mix a fraction of Hartree-Fock (HF) exchange with local or semilocal exchange-correlation functionals also show serious discrepancies. Yang and co-workers \cite{Yang14}, Lordi \cite{Lordi13}, and Lindstr\"{o}m \emph{et al.} \cite{Lindstrom16} have found a negative-U behavior in (Te$_\text{Cd}$). In contrast, Biswas and Du \cite{Biswas12} have pointed out that $(\text{Te}_\text{Cd})$ is a deep donor with (+2/+) and (+/0) transition levels at VBM + 0.38 eV and VBM + 0.58 eV, respectively.

 In order to investigate these large discrepancies among theoretical calculations, in the present work we investigate the formation energies, charge transition levels and quasiparticle defect states of $(\text{Te}_\text{Cd})$ in CdTe using the state-of-the-art DFT$\hspace{0.05cm}+\hspace{0.05cm}$\emph{GW} formalism \cite{Hedstrom06,Rinke09,Malashevich14,Flores15_1}, which is free of the well-known band gap error of DFT. According to our results, $(\text{Te}_\text{Cd})$ induces a deep level at VBM + 0.99 eV, exhibiting a negative-U effect. Moreover, the optical excitation of the $(\text{Te}_\text{Cd})^0$ configuration to the positively charged state, followed by the capture of an electron from the conduction bands is consistent with the 1.1-eV center observed in both absorption \cite{Davis93} and photoluminescence (PL) \cite{Bryant68} measurements at cryogenic temperatures.

\section{Methods}
\vspace{0.1cm}
\subsection{Computational methods}
\vspace{0.3cm}
Our DFT calculations were performed using the Quantum-ESPRESSO code \cite{Giannozzi2009}. Electron-ion interactions were described by GBRV ultrasoft pseudopotentials \cite{Garrity2014}, whereas the generalized gradient approximation to the exchange and correlation functional of Perdew, Burke, and Ernzerhof (PBE) \cite{Perdew96} was employed. A kinetic energy cutoff of 36 Ry for the plane-wave basis set expansion and 200 Ry to represent the charge density were used. To avoid finite-size effects as much as possible, the defect calculations were performed within large 512-atom cubic supercells. The atomic structures were relaxed until the Hellmann-Feynman forces were less than 0.001 Ry/bohr. The \textbf{k}-point sampling was restricted to the $\Gamma$ point.

Many-body $G_0W_0$ calculations with defect supercells were performed using the WEST code \cite{Pham13,Govoni15}, which avoids an explicit sum over empty orbitals by using a technique called projective eigendecomposition of the dielectric screening (PDEP) \cite{Pham13}, evaluating the correlation self-energy by a Lanczos-chain algorithm \cite{Rocca08}. In our calculations we used 200 projective dielectric eigenpotential basis vectors to represent the inverse of the Hermitian dielectric matrix and 30 Lanczos steps to evaluate the irreducible polarizability. Our tests show that these parameters are sufficient to obtain a well-converged band gap within 0.1 eV. For the absolute position of the VBM we used $\Delta E_\text{VBM} = -0.74$ eV as obtained in Ref. \cite{Gruneis14} employing the \emph{GW}$\Gamma$ approximation, that includes a first-order vertex correction in the self-energy and the effect of spin-orbit coupling. Optimized norm-conserving Vanderbilt pseudopotentials (ONCV) \cite{Hamann13} with 20 and 16 valence electrons for Cd and Te atoms, respectively, and a plane-wave energy cutoff of 70 Ry were employed. A considerable improvement in computational efficiency was obtained employing ONCV pseudopotentials, as the plane-wave cutoff requirements with semicore states are modest compared to the conventional Kleinman-Bylander \cite{Kleinman82} representation.

The $G_0W_0$ band gap of bulk CdTe is calculated to be 1.56 eV, in excellent agreement with the room temperature band gap of 1.5 eV, as well as with previous calculations \cite{Klimes14}. Quasiparticle corrections to Kohn-Sham (KS) eigenvalues were obtained using 64-atom supercells at the $\Gamma$ point only. These corrections were then applied to the KS eigenvalues obtained from DFT calculations employing 512-atom supercells.

\vspace{0.1cm}
\subsection{Defect formation energies}
\vspace{0.3cm}

The formation energy of a defect in charge state $q$ and arbitrary ionic configuration \textbf R can be expressed as \cite{Jain11}
\begin{eqnarray}
\hspace{1.5cm}E^f_q[\textbf{R}] = E_q[\textbf{R}] - E_\text{ref} + qE_F,
\end{eqnarray}
\begin{eqnarray}
\hspace{1.5cm}E_\text{ref} \equiv E^\text{CdTe}_\text{bulk} + \sum_i n_i\mu_i,
\end{eqnarray}
where $E_q[\textbf{R}]$ is the total energy of the system in charge state $q$ and atomic positions $\textbf{R}$, and $E_\text{ref}$ is the energy of a reference system with the same number of atoms as the supercell containing an isolated defect. The integer $n_i$ indicates the number of $i$ elements (Cd or Te) that have been added ($n_i > 0$) or removed ($n_i < 0$) from the supercell, and $\mu_i$ is the chemical potential of the element $i$, and $E_F$ is Fermi energy.

The chemical potentials are defined by the experimental growth conditions. For the case of CdTe, the Cd-rich limit is defined by imposing an equilibrium between the system and a reservoir of bulk Cd, whereas for the Te-rich limit $\mu_\text{Te} $ is equivalent to the energy of bulk Te. Therefore, $\mu_{\text{Cd}}$ and $\mu_{\text{Te}}$ are assumed under Cd-rich conditions to be $\mu_\text{Cd} $ = $\mu_\text{Cd (bulk)} $ and $\mu_\text{Te} $ = $\mu_\text{CdTe} - \mu_\text{Cd}$. Similarly, under Te-rich conditions, $\mu_\text{Te} $ = $\mu_\text{Te (bulk)} $ and $\mu_\text{Cd} $ = $\mu_\text{CdTe} - \mu_\text{Te}$. The stability condition for CdTe requires $E^f[\text{CdTe}] < \Delta\mu_{\text{Te}} < 0$, and $E^f[\text{CdTe}] < \Delta\mu_{\text{Cd}} < 0$, where $E^f[\text{CdTe}] $ is the formation energy of CdTe, which is calculated to be $-0.91$ eV, in good agreement with the experimental value of $-0.96$ eV \cite{Haynes14}, and $\Delta\mu_{\text{i}}$ is the relative chemical potential referenced to their respective reservoirs, e.g., $\Delta\mu_{\text{Te}} = \mu_\text{Te}- \mu_\text{Te (bulk)}$.
\vspace{0.1cm}
\subsection{DFT$\hspace{0.05cm}+\hspace{0.05cm}$GW formalism}
\vspace{0.3cm}
The formation energy of a defect in charge state $q\hspace{-0.05cm}-\hspace{-0.07cm}1$ is given by
\begin{eqnarray}
E^f_{q-1}[\textbf{R}_{q-1}] = E_{q-1}[\textbf{R}_{q-1}] - E_\text{ref} + (q-1)E_F.
\end{eqnarray}
By adding and substracting first $E_{q-1}[\textbf{R}_{q}]$ and then $E_{q}[\textbf{R}_{q}]$, we have \cite{Flores15_1}
\begin{equation}
\begin{split}
E^f_{q-1}[\textbf{R}_{q-1}] &= \left \{ E_{q-1}[\textbf{R}_{q}] - E_{q}[\textbf{R}_{q}] \right \} \\
&+ \left \{ E_{q-1}[\textbf{R}_{q-1}] - E_{q-1}[\textbf{R}_{q}]\right \} \\
&+ E^f_{q}[\textbf{R}_{q}] - E_F \\
&\equiv E_\text{QP} + E_{\text{relax}} + E^f_{q}[\textbf{R}_{q}] - E_F,
\end{split}
\end{equation}
where $\textbf{R}_{q}$ corresponds to the minimum energy configuration for the charge state $q$. The first term is a quasiparticle energy (i.e., an electron addition or removal energy) and may be calculated using the many-body perturbation theory based on the $GW$ approximation \cite{Hedin65, Hybertsen85}. The second term corresponds to a relaxation energy and may be evaluated at DFT level, since we only calculate energy differences between configurations with the same number of electrons.

Using Kohn-Sham wave functions $\psi_{n,k}^{\text{KS}} $ and energies $\epsilon_{n,k}^{\text{KS}} $ as mean-field starting points for the construction of G and W ($G_0W_0$ approximation), we calculate the quasiparticle energies $E^\text{QP}_{n,k}$ within a first-order perturbation theory approximation as

\begin{equation}
E^\text{QP}_{n,k} = \epsilon_{n,k}^{\text{KS}}  + \left< \psi_{n,k}^{\text{KS}} |\Sigma (E^{\text{QP}}_{n,k}) -  V_\text{xc}| \psi_{n,k}^{\text{KS}}\right>,
\end{equation}
which comes from replacing the KS exchange-correlation potential $V_{xc}$ with the self-energy operator $\Sigma$. When the reference state is an open-shell system, wave functions and energies from spin-polarized DFT calculations were used as mean-field starting points.

Considering the computational demands, we employed a cubic 64-atom supercell to calculate the quasiparticle corrections to the DFT eigenvalues at the $\Gamma$ point only. These corrections were then applied to the KS eigenvalues of 512-atom supercells to obtain the quasiparticle energies referenced to the average electrostatic potential of bulk CdTe. This approach is justified because we consider finite-size effects at the DFT level. Moreover, quasiparticle corrections are largely invariant with respect to the supercell size \cite{Flores15_1,Choi09,Chen13} and, at the high-symmetry points their differences are up to 0.1 eV.  The relaxation energies were calculated using 512-atom supercells.

\section{Results and discussion}

   Starting from the ground state configuration $(\text{Te}_\text{Cd})^{+2}$, we can obtain the formation energies for different charge states using Eq. (4). A key observation is that the self-interaction error will mostly cancel in the first difference of Eq. (1), since it has all the valence bands full and all the conduction bands empty.

     We should note that the absolute position of the VBM of bulk CdTe obtained using the PBE exchange-correlation functional was corrected by $\Delta E_\text{VBM} = -0.74$  eV. Hence, the energy change due to the exchange of electrons and holes with the carrier reservoirs for the $(\text{Te}_\text{Cd})^{+2}$ configuration differs by $+2 \times \Delta E_\text{VBM} = -1.48 $ eV, as compared to PBE. Moreover, in the case of CdTe, the widely used screened hybrid functional of Heyd, Scuseria and Ernzerhof (HSE) \cite{Heyd05} only partially corrects the self-interaction error, lowering the energy of the VBM by 0.51 eV with respect to PBE \cite{Gruneis14}, resulting in a formation energy $0.46$ eV higher than our results (an illustrative comparison between LDA and HSE06 can be found in Fig. 4 of Ref. \cite{Lindstrom16}). Du \cite{Du15} has recently stressed the importance of the correct absolute positions of VBM and CBM for reliable predictions of charge transition levels. It deserves noting that corrections for electron and chemical reservoirs have been recently proposed \cite{Freysoldt16}.

\begin{figure}[h]%
 \centering
 \vspace{0.4cm}
 \includegraphics[width=3.5cm]{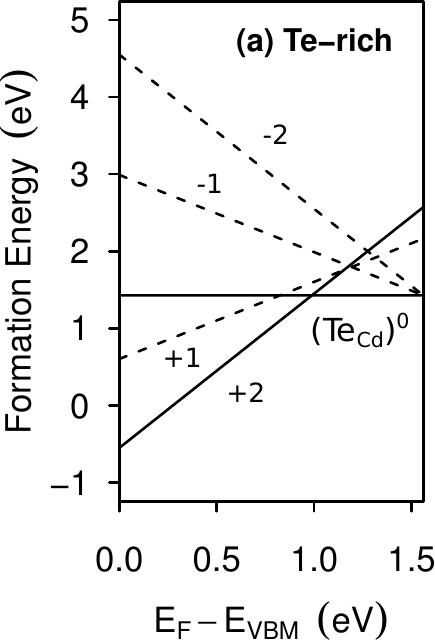}
  \label{tab:f1}
 \hspace{0.5cm}
 \includegraphics[width=3.5cm]{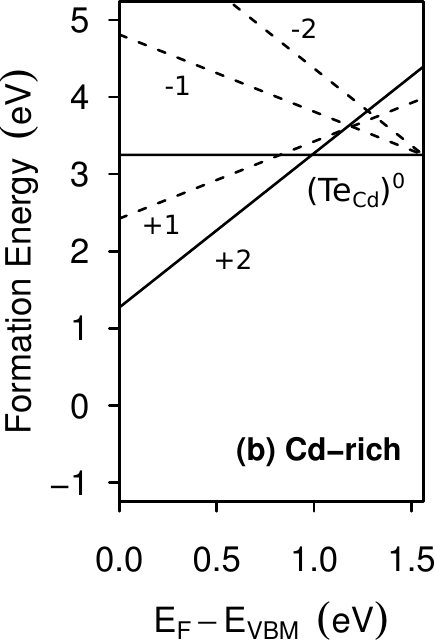}
  \caption{Calculated formation energies of $(\text{Te}_\text{Cd})$ in various charge states as a function of the Fermi level inside the band gap. The stable charge states are shown by solid lines.}
\end{figure}

   The calculated defect formation energies are plotted as a function of the Fermi level in Figure 1. Table I shows the contributions to the formation energies coming from quasiparticle and relaxation energies according to Eq. (4). The formation energy of $(\text{Te}_\text{Cd})$ in the neutral charge state is found to be 1.45 eV for the Te-rich limit, and 3.27 eV for the Cd-rich limit. Our results indicate that $(\text{Te}_\text{Cd})$ exhibits a negative-U behavior that causes the $(+1)$ charge state to be unstable. The $(+2/0)$ charge transition level is found to be deep in the band gap, at VBM + 0.99 eV. For low values of the Fermi energy, the Te antisite will be in a double positive charge state, whereas for $n$-type CdTe, the neutral charge state will be favored.

\begin{table}[h]
\centering
\vspace{0.2cm}
\begin{tabular*}{0.60\textwidth}{@{\extracolsep{\fill}}ccc}
\hline \hline
\T & $E_\text{QP} - E_\text{VBM}$ (eV) & $E_{\text{relax}}$ (eV)\B\\
\hline
\T \T $E^f_{+1}$ & $1.53$ & $-0.21\hspace{0.25cm}$ \\
\T $E^f_{0}$ & $1.05$ & $- 0.23\hspace{0.25cm}$\\
\T $E^f_{-1}$ & $1.56$ & $0.01$\\
\T $E^f_{-2}$ & $1.56$ & $0.00$\B\\
\hline \hline
\end{tabular*}
\caption{\label{tab:table1}Contributions to the formation energies of $(\text{Te}_\text{Cd})$ coming from quasiparticle and relaxation energies, according to Eq. (4).}
\end{table}

\begin{figure}[h]%
 \centering
 \vspace{0.1cm}
 (a)\hspace{-0.4cm}
 \includegraphics[width=3.3cm]{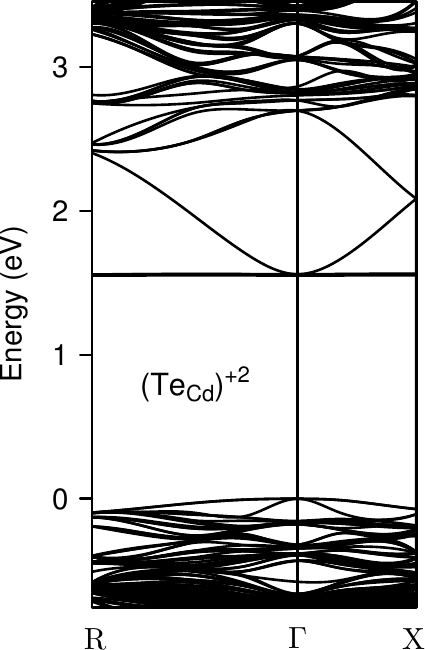}
 \hspace{0.7cm}
 (b)\hspace{-0.4cm}\includegraphics[width=3.3cm]{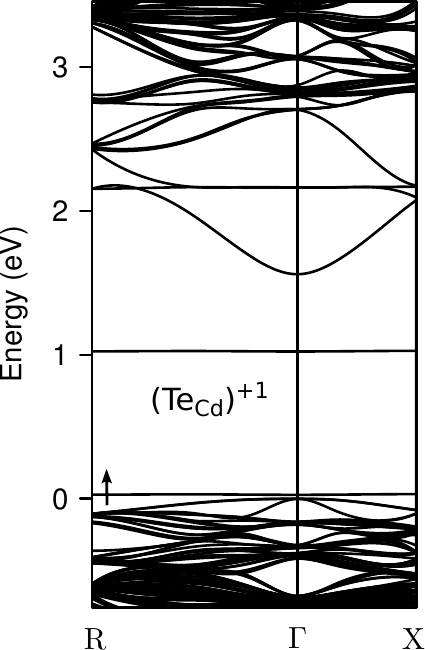}
 \caption{Theoretical band structure of (a) $(\text{Te}_\text{Cd})^{+2}$, and (b) $(\text{Te}_\text{Cd})^{+1}$, calculated by using 512-atom supercells. In (b), the arrow indicates the occupation of the energy level in the band gap.}
 \label{f2}
\end{figure}

    Figure 2 shows the electronic band structures of $(\text{Te}_\text{Cd})$ in (+1), and (+2) charge states calculated by using large 512-atom supercells. A scissors operator, consisting in a shift to the defect level and a rigid shift to the conduction bands so as to recover the $G_0W_0$ quasiparticle band gap, was applied to correct the KS band structure. In the ideal $T_d$ symmetry, the Te antisite induces a triple-degenerate energy level inside the band gap, and it would be unstable with respect to symmetry-lowering distortions that minimize the total electronic energy. However, the ground state configuration $(\text{Te}_\text{Cd})^{+2}$ maintains the $T_d$ symmetry, because the triple-degenerate energy level is unoccupied [Figure 2 (a)]. On the other hand, $(\text{Te}_\text{Cd})^{+1}$ [Figure 2 (b)] and $(\text{Te}_\text{Cd})^{0}$ [Figure 3] undergo static Jahn-Teller distortions \cite{Opik57}. Two A$_1$ and one E double-degenerate level can be identified (labeled $u, v$, and $e$, respectively). A $T_d$ to $C_{3v}$ distortion gives a $u^2v^2e$ ground state configuration, where $u$ is located below the VBM, $v$ remains isolated in the band gap, and the double-degenerate level E is resonant with the conduction bands.

\begin{figure}[h]%
 \centering
 \vspace{0.3cm}
 \includegraphics[width=6.55cm]{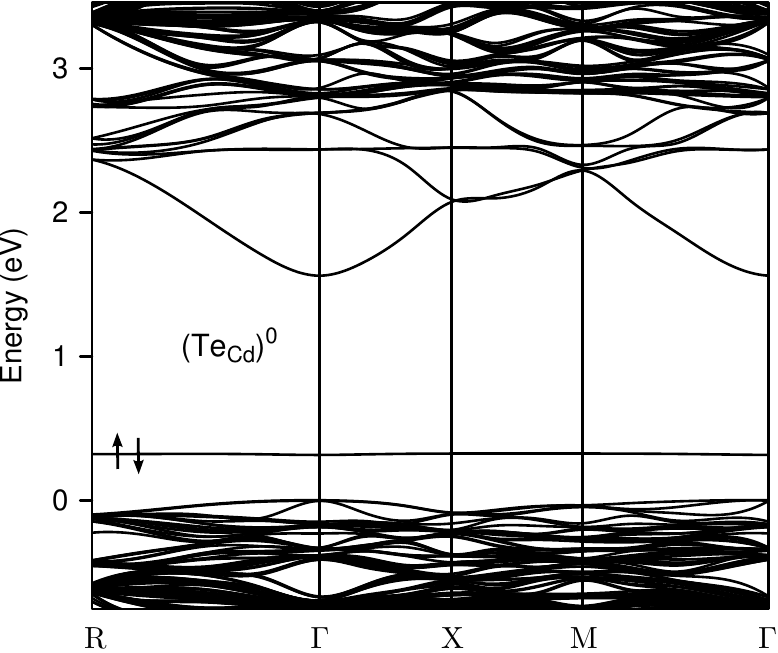}
 \hspace{0.1cm}
 \includegraphics[width=3.6cm]{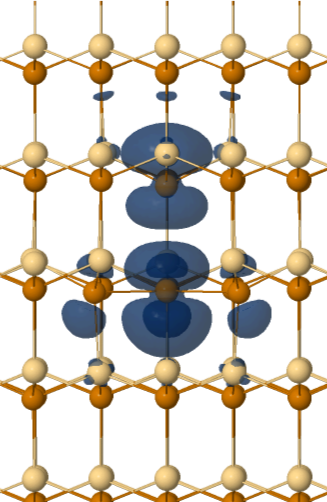}
 \caption{(Color online) Theoretical band structure and charge density isosurface ($\rho =$ 0.0005 $e$/Bohr$^3$) of the energy level in the band gap of (Te$_\text{Cd}$)$^0$, calculated by using a 512-atom supercell. Dark spheres are Te atoms and light spheres are Cd atoms. The crystal is oriented along the $\langle 111 \rangle$ direction.}
\label{f3}
\end{figure}
    The ground state $(\text{Te}_\text{Cd})^{+2}$ configuration has an empty triple-degenerate energy level very close to the CBM, as shown in Figure 2 (a). The addition of one electron induces a Jahn-Teller distortion, lifting the degeneracy. As the system has now a partially occupied highest energy level [Figure 2 (b)], it is expected to increase its energy if an additional electron is captured, due to the Coulombic repulsion. However, the presence of a second electron induces an energy-lowering structural distortion that supply a net effective attractive interaction (negative-U effect) that overcome Coulombic repulsion. Therefore, electrons are likely to be trapped by pairs at the defect. Our calculated value of U = $\epsilon (+|0) - \epsilon (+2|+) $ is found to be $-0.38$ eV.

 According to Figure 1, in $p$-type conditions, the Te antisite is favorable to be in a double positive charge state. It should tend to transfer its electrons to uncompensated acceptors such as Cd vacancies, which are present at significant concentrations in CdTe \cite{Szeles04,Shepidchenko15}. Although the $(+2/0)$ level is deep in the band gap, the unoccupied triple-degenerate energy level close to the CBM may easily capture a pair of electrons from the conduction bands. If so, a $T_d$ to $C_{3v}$ Jahn-Teller distortion would lift the degeneracy, leaving a fully occupied isolated energy level in the band gap at VBM + 0.3 eV [Figure 3].

\begin{figure}[h]%
 \centering
 \vspace{0.3cm}
 \includegraphics[width=7.5cm]{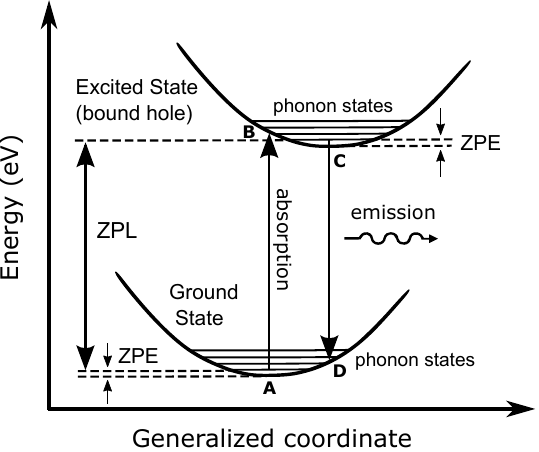}
 \caption{Configuration-coordinate diagram for the excitation cycle of $(\text{Te}_\text{Cd})^{0}$ .}
\label{fig4}
\vspace{0.3cm}
\end{figure}

 As noted above, the Te antisite is energetically favorable to be in the neutral charge state in $n$-type CdTe. However, this configuration may also be metastable when the position of the Fermi level is near the middle of the gap. Moreover, the optical excitation of $(\text{Te}_\text{Cd})^{0}$ to the positively charged state, followed by the capture of an electron from the conduction bands is consistent with the observed absorption peak near 1.1 eV \cite{Davis93,Simonds06}, as well as with the 1.1-eV band generally found in PL measurements \cite{Bryant68,Bowman88,Krustok96,Zazvorka16}. The former was attributed to localized defect states within the band gap \cite{Davis93}, whereas the latter has been associated with donor-acceptor pair (DAP) transitions \cite{Krustok96}. More recently, it was proposed that the PL band could be caused by a transition from an excited state activated by carrier capture (component 9 in Ref. \cite{Zazvorka16}).

  To gain further understanding on this issue, we calculate the energy of the zero-phonon line (ZPL), which allows us to compare our calculations to experimental results at low temperatures. The zero-point vibration states will raise the energies of the ground state and excited configurations by a value of the order of a few tens meV, called zero-point energy (ZPE). The difference between the ZPE of the ground state and excited configurations is expected to be even smaller, of the order on a a few meV. Therefore, the ZPL can be well approximated by the sum of the excitation energy for promoting one electron from the localized energy level of $(\text{Te}_\text{Cd})^{0}$ (the gap-state in Fig. 3) to the conduction bands (transition A $\rightarrow$ B in Fig. 4), and the subsequent relaxation energy of the excited configuration (transition B $\rightarrow$ C in Fig. 4); the latter produces a shift in the absorption energy (a Stoke shift). We use constrained DFT \cite{Dederichs84} to calculate the Stokes shift. This method allows one to define constraints on the charge density, and has been successfully applied to Nitrogen-Vacancy \cite{Gali09_1,Gali09,Choi12}, and Silicon-Vacancy \cite{Gali13} color centers in diamond. The expected error in this approach is small, as the Stokes shift corresponds to the energy difference between two different ionic configurations with the same electronic configuration \cite{Gali09}; the same principle is used in the DFT$\hspace{0.05cm}+\hspace{0.05cm}$\emph{GW} formalism \cite{Flores15_1}.

   Our calculated energies for the vertical absorption (A $\rightarrow$ B) and the Stokes shift (B $\rightarrow$ C) are 1.26 eV and $-0.14$ eV, respectively; thus, the ZPL is calculated to be 1.12 eV. This result agrees well with the 1.1-eV center observed in both absorption \cite{Davis93} and emission \cite{Davis93,Bryant68} at cryogenic temperatures.

\begin{figure*}[!ht]%
\vspace{0.4cm}
 \centering
 (a)\hspace{-0.35cm}\includegraphics[width=3.2cm]{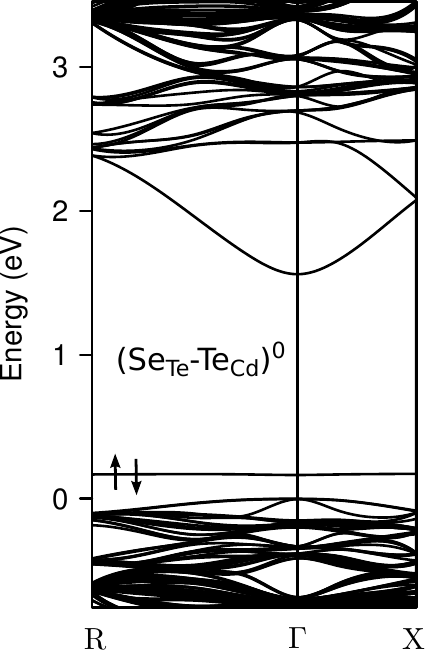}
 \hspace{-0.2cm}
 \includegraphics[width=2.76cm]{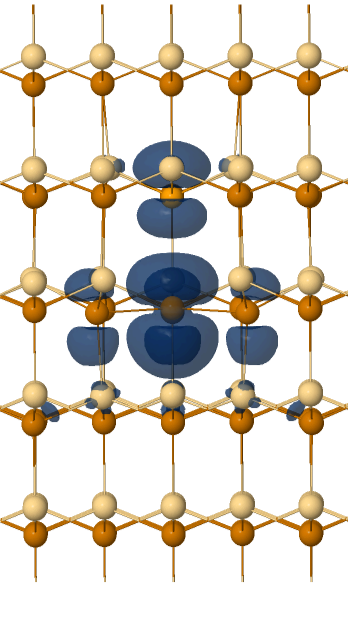}
 \hspace{0.5cm}
 (b)\hspace{-0.35cm}\includegraphics[width=3.2cm]{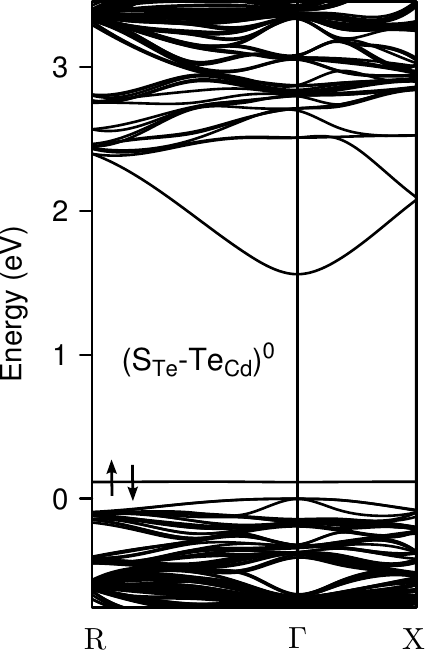}
 \hspace{-0.2cm}
 \includegraphics[width=2.76cm]{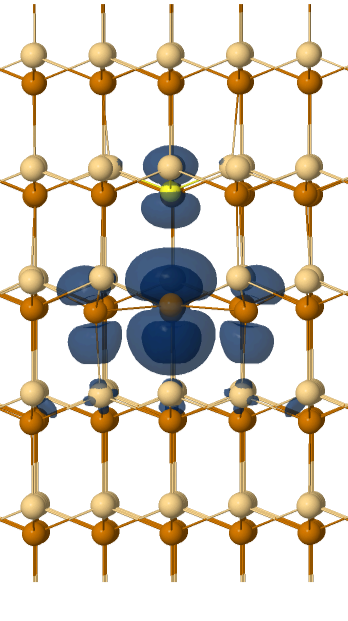}
 \vspace{0.5cm}\\
 (c)\hspace{-0.35cm}\includegraphics[width=3.2cm]{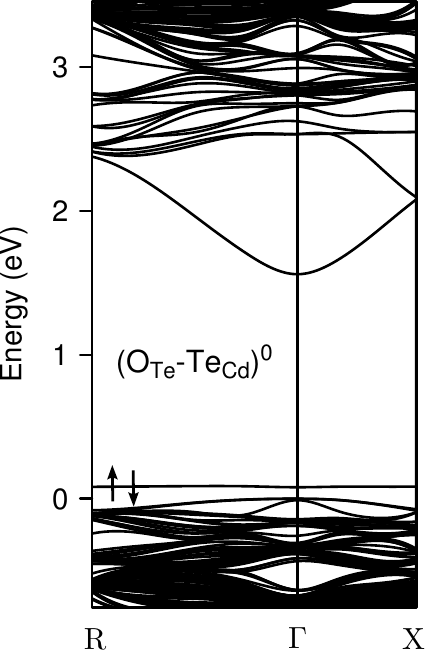}
 \hspace{-0.2cm}
 \includegraphics[width=2.74cm]{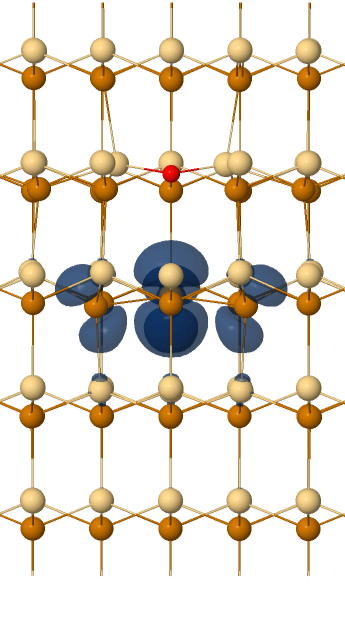}
 \caption{(Color online) Theoretical band structure and charge density isosurfaces ($\rho =$ 0.0005 $e$/Bohr$^3$) of the energy levels in the band gap for the neutral complexes: (a) $(\text{Se}_{\text{Te}}\hspace{-0.05cm}-\hspace{-0.05cm}\text{Te}_\text{Cd})^0$, (b) $(\text{S}_{\text{Te}}\hspace{-0.05cm}-\hspace{-0.05cm}\text{Te}_\text{Cd})^0$, and (c) $(\text{O}_{\text{Te}}\hspace{-0.05cm}-\hspace{-0.05cm}\text{Te}_\text{Cd})^0$, calculated by using 512-atom supercells. The arrows indicate the occupation of the energy level in the band gap. Dark spheres are Te atoms and light spheres are Cd atoms. The crystals are oriented along the $\langle 111 \rangle$ direction.}
 \vspace{0.1cm}
\end{figure*}

 Having identified the Te antisite in the neutral charge state as hole trap, we should note that in order to limit any potential deleterious impact to carrier transport, Te-poor grown conditions are desirable. However, most polycrystalline CdTe films require high growth temperatures resulting in a Te-excess material (due to the lost of Cd during the growth process) \cite{Yujie03,Moure-Flores12}. To solve this problem, it was recently proposed that the incorporation of oxygen passivates the gap states associated with $(\text{Te}_\text{Cd})^0$, by forming $(\text{O}_\text{Te}-\text{Te}_\text{Cd})$ complexes \cite{Flores15_1}.

 To investigate more deeply the beneficial effects of oxygen incorporation, we perform DFT calculations considering three distinct isovalent impurities: selenium, sulfur, and oxygen. For the cases of selenium [Figure 5 (a)] and sulfur [Figure 5 (b)], the electronegativity of the impurity atom is reflected in the size of its antibonding molecular orbital; as consequence, the energy of the $(\text{Te}_\text{Cd})^0$ gap state, located at VBM + 0.3 eV, is lowered by 0.13 eV and 0.18 eV, respectively. Remarkably, isovalent oxygen completely removes the antibonding interaction along the $C_{3v}$ rotation axis, lowering the position of the gap state by $0.24$ eV [Figure 5 (c)]. In the case of $(\text{O}_\text{Te}-\text{Te}_\text{Cd})$, the localized energy level is located at VBM + 0.06 eV; thus, hole trapping is unlikely to occur.

\section{Conclusions}
In summary, we have investigated the formation energies, charge transition levels and quasiparticle defect states of the Te antisite in CdTe within the DFT$\hspace{0.05cm}+\hspace{0.05cm}$\emph{GW} formalism. We find that $(\text{Te}_\text{Cd})$ is a negative-U defect, inducing a deep donor level at VBM + 0.99 eV. In addition, our results suggest that the $\sim$1.1 eV band, visible in both luminescence and absorption experiments can be associate with the $(\text{Te}_\text{Cd})^0$ defect, which acts as a hole trap.

\section*{Acknowledgment}
This work was supported by the FONDECYT Grant No. 1130437. Powered@NLHPC: This research was partially supported by the supercomputing infrastructure of the NLHPC (ECM-02).


 \bibliography{Te_Cd}


\end{document}